\documentclass[12pt]{article}
\usepackage{amsmath}
\usepackage{amssymb}
\usepackage{epsfig}
\usepackage{euscript}
\def\mathswitchr#1{\relax\ifmmode{\mathrm{#1}}\else$\mathrm{#1}$\fi}

\newcommand {\pslash}{\hbox{$\not\hbox{\kern-2.3pt $p$}$}}

\newcommand{\FYFS}{F_{\mathrm{YFS}}}
\usepackage{cite}

\usepackage{epic}

\def\alf1{ {\alpha\over\pi} }
\def\rQCED{{\rm QCED}}
\begin{document}
\begin{titlepage}
\begin{center}
{\bf \large Phenomenology of the Interplay between IR-Improved DGLAP-CS Theory and NLO ME Matched Parton Shower MC Precision}\\
\vspace{2mm}
    S.K. Majhi
 \footnote{Work supported by 
grant Pool No. 8545-A, CSIR, IN.}\\
      Indian Association for the Cultivation of Science, Kolkata, India\\
        E-mail: tpskm@iacs.res.in\\
A. Mukhopadhyay\\%
      Baylor University, Waco, TX, USA\\
        E-mail: aditi\_mukhopadhyay@baylor.edu\\
B.F.L. Ward\\%
      Baylor University, Waco, TX, USA\\
        E-mail: bfl\_ward@baylor.edu\\
S.A. Yost
 \footnote{Work supported in part by U.S.
D.o.E. grant DE-FG02-10ER41694 and grants from The Citadel Foundation.}\\
      The Citadel, Charleston, SC, USA\\
        E-mail: scott.yost@citadel.edu\\
\end{center}
\vspace{2mm}
\centerline{\bf Abstract}
We present the current status of the application of our approach of {\it exact} amplitude-based resummation in quantum field theory to precision QCD calculations, by realistic MC event generator methods, as needed for precision LHC physics. In this ongoing program of research, we discuss recent results as they relate to the interplay of the attendant IR-Improved DGLAP-CS theory of one of us and the precision of exact NLO matrix element matched parton shower MC's in the Herwig6.5 environment in relation to recent
LHC experimental observations. There continues to be reason for optimism in the attendant comparison of theory and experiment.\\
\vspace{1cm}
\begin{center}
 BU-HEPP-12-02, Aug., 2012\\
\end{center}

\end{titlepage}
%

 
\def\Kmax{K_{\rm max}}\def\ieps{{i\epsilon}}\def\rQCD{{\rm QCD}}

\section{\bf Introduction}\par
With the recent announcement~\cite{atlas-cms-2012} 
of an Englert-Brout-Higgs (EBH)~\cite{EBH} candidate 
boson after the start-up and successful running 
of the LHC for 2.5 years, the era of precision QCD, 
by which we mean
predictions for QCD processes at the total precision tag of $1\%$ or better,
is squarely upon us. The attendant need for exact, amplitude-based 
resummation of large higher order effects
is now more paramount, given the expected role of precision comparison between theory and experiment in determining the detailed properties of the newly discovered EBH boson candidate. Three of us(B.F.L.W,S.K.M,S.A.Y) have argued elsewhere~\cite{radcor2011,qced} that
such resummation allows one to have better than 1\% theoretical precision 
as a realistic goal in such comparisons, so that one can indeed 
distinguish new physics(NP) from higher order SM processes and can distinguish 
different models of new physics from one another as well. 
In what follows, we present the status of this approach to precision QCD for the LHC in connection with its attendant IR-improved DGLAP-CS~\cite{dglap,cs} theory~\cite{irdglap1,irdglap2} realization via HERWIRI1.031~\cite{herwiri} 
in the HERWIG6.5~\cite{herwig} environment
in interplay with NLO exact, matrix element matched parton shower MC precision issues. We will employ the MC@NLO~\cite{mcatnlo} methodology to realize
the attendant exact, NLO matrix element matched parton shower MC
realizations for both HERWIRI1.031 and HERWIG6.5 in our corresponding
comparisons with recent LHC data that we present herein.
\par
The discussion will therefore be seen to continue the strategy of building on existing platforms to develop and realize a path toward precision QCD for the physics of the LHC. We exhibit explicitly a union of the new IR-improved DGLAP-CS theory and the MC@NLO realization of exact NLO matrix element(ME) matched parton shower MC theory. As our ultimate goal is a provable precision tag on our theoretical predictions, we note that we are also pursuing the implementation~\cite{elswh} of the new IR-improved 
DGLAP-CS theory for
HERWIG++~\cite{hwg++}, HERWIRI++,
for PYTHIA8~\cite{pyth8} and for SHERPA~\cite{shrpa}, as well as
the corresponding NLO ME/parton shower matching realizations in the POWHEG~\cite{pwhg} framework. For, one of the strongest cross checks on theoretical precision is the difference between two independent realizations of the attendant theoretical calculation. Such cross checks will appear elsewhere~\cite{elswh}.
\par
In order to expose properly the interplay between the NLO ME matched parton shower MC precision and the new IR-improved DGLAP-CS theory, we set the stage in the next section by showing how the latter theory follows 
naturally in the effort to obtain
a provable precision from our approach~\cite{qced} to precision LHC physics. 
In the interest of completeness, we review this latter approach, 
which is an amplitude-based QED$\otimes$QCD($\equiv\text{QCD}\otimes\text{QED}$) exact resummation 
theory~\cite{qced} 
realized by MC methods, in the next section as well. We then turn in Section 3 to the applications to the recent data on single heavy gauge boson production at the LHC from the perspective of the analysis in Refs.~\cite{herwiri} of the analogous processes at the Tevatron, where we will focus in this Letter on the single Z/$\gamma*$ production and decay to lepton pairs for definiteness. The other heavy gauge boson processes will be taken up elsewhere~\cite{elswh}. Section 4 contains our summary remarks.
\par
\section{Recapitulation}

The starting point for what we discuss here may be taken as the 
fully differential representation
\begin{equation}
d\sigma =\sum_{i,j}\int dx_1dx_2F_i(x_1)F_j(x_2)d\hat\sigma_{\text{res}}(x_1x_2s)
\label{bscfrla}
\end{equation}
of a hard LHC scattering process
using a standard notation so that the $\{F_j\}$ and 
$d\hat\sigma_{\text{res}}$ are the respective parton densities and 
reduced hard differential cross section where we indicate the that latter 
has been resummed
for all large EW and QCD higher order corrections in a manner consistent
with achieving a total precision tag of 1\% or better for the total 
theoretical precision of (\ref{bscfrla}). The key theoretical issue
for precision
QCD for the LHC is then the proof of the correctness of the value of the 
total theoretical precision $\Delta\sigma_{\text{th}}$ of (\ref{bscfrla}). 
This precision can be represented as follows:
\begin{equation}
\Delta\sigma_{\text{th}}= \Delta F \oplus\Delta\hat\sigma_{\text{res}}
\label{eqdecomp1}
\end{equation}
where $\Delta A$ is the contribution of the uncertainty
on the quantity $A$ to $\Delta\sigma_{\text{th}}$\footnote{Here, we discuss the situation in which the two errors in (\ref{eqdecomp1}) are independent
for definiteness; (\ref{eqdecomp1}) has to be modified accordingly when
they are not.}.
In order to validate the  application of a given 
theoretical prediction to precision 
experimental observations, for the discussion of the signals and the
backgrounds for 
both Standard Model(SM) and new physics (NP) studies, and more specifically
for the overall normalization
of the cross sections in such studies, the
proof of the correctness of the value of the 
total theoretical precision $\Delta\sigma_{\text{th}}$ 
is essential. If a calculation
with an unknown value of $\Delta\sigma_{\text{th}}$ 
is used for the attendant studies, the NP can be missed.
This point simply cannot be emphasized too much.\par
In the interest of completeness here,
we note that, by our definition, $\Delta\sigma_{\text{th}}$ is the 
total theoretical uncertainty that comes from the physical precision 
contribution and the technical precision contribution~\cite{jadach-prec}:
the physical precision contribution, $\Delta\sigma^{\text{phys}}_{\text{th}}$,
arises from such sources as missing graphs, approximations to graphs, 
truncations, etc.; the technical precision contribution, 
$\Delta\sigma^{\text{tech}}_{\text{th}}$, arises from such sources as 
bugs in codes\footnote{We have in mind that all gross errors such as those that give obviously wrong results, as determined by cross checks, are eliminated and we have left programming errors such as those in the logic: suppose for programming error reasons a DO-loop ends at 999 steps instead of the intended 1000 steps, resulting in a per mille level error, that could alternate in sign from event to event. As per mille level accuracy is good enough in many applications, the program would remain reliable, but it would have what we call a technical precision error at the per mille level.}, numerical rounding errors,
convergence issues, etc. The total theoretical error is then 
given by
\begin{equation}
\Delta\sigma_{\text{th}}=\Delta\sigma^{\text{phys}}_{\text{th}}\oplus \Delta\sigma^{\text{tech}}_{\text{th}}.
\end{equation}
The desired value for $\Delta\sigma_{\text{th}}$, which 
depends on the  specific
requirements of the observations, as a general rule, should fulfill
$\Delta\sigma_{\text{th}}\leq f\Delta\sigma_{\text{expt}}$, 
where $\Delta\sigma_{\text{expt}}$ is the respective experimental error
and $f\lesssim \frac{1}{2}$. This would assure that
the theoretical uncertainty does not significantly adversely affect the 
analysis of the data for physics studies.
\par

In order to realize such precision in a provable way, we have 
developed the $\text{QCD}\otimes\text{QED}$ resummation theory in Refs.~\cite{qced}
for the reduced cross section in (\ref{bscfrla}) and for the
resummation of the evolution of the parton densities therein as well.
In the interest of completeness and also because the theory in
Refs.~\cite{qced} is not widely known, we recapitulate it here briefly.
Specifically, for both the resummation of the reduced cross section
and that of the evolution of the parton densities, the master formula 
may be identified as
\begin{eqnarray}
&d\bar\sigma_{\rm res} = e^{\rm SUM_{IR}(QCED)}
   \sum_{{n,m}=0}^\infty\frac{1}{n!m!}\int\prod_{j_1=1}^n\frac{d^3k_{j_1}}{k_{j_1}} \cr
&\prod_{j_2=1}^m\frac{d^3{k'}_{j_2}}{{k'}_{j_2}}
\int\frac{d^4y}{(2\pi)^4}e^{iy\cdot(p_1+q_1-p_2-q_2-\sum k_{j_1}-\sum {k'}_{j_2})+
D_\rQCED} \cr
&\tilde{\bar\beta}_{n,m}(k_1,\ldots,k_n;k'_1,\ldots,k'_m)\frac{d^3p_2}{p_2^{\,0}}\frac{d^3q_2}{q_2^{\,0}},
\label{subp15b}
\end{eqnarray}\noindent
where $d\bar\sigma_{\rm res}$ is either the reduced cross section
$d\hat\sigma_{\rm res}$ or the differential rate associated to a
DGLAP-CS~\cite{dglap,cs} kernel involved in the evolution of the $\{F_j\}$ and 
where the {\em new} (YFS-style~\cite{yfs}) {\em non-Abelian} residuals 
$\tilde{\bar\beta}_{n,m}(k_1,\ldots,k_n;k'_1,\ldots,k'_m)$ have $n$ hard gluons and $m$ hard photons and we show the final state with two hard final
partons with momenta $p_2,\; q_2$ specified for a generic $2f$ final state for
definiteness. The infrared functions ${\rm SUM_{IR}(QCED)},\; D_\rQCED\; $
are defined in Refs.~\cite{qced,irdglap1,irdglap2}. This  
simultaneous resummation of QED and QCD large IR effects is exact.\par 
The key components in the master formula (\ref{subp15b}) have the following physical meanings. The exponent ${\rm SUM_{IR}(QCED)}$ sums up to the infinite  order the maximal leading IR singular terms in the cross section in the language of
Ref.~\cite{gatheral} for soft emission 
below a dummy parameter $K_{\text{max}}$ and the exponent
$D_\rQCED$ does the same for the regime above $K_{\text{max}}$ so that
(\ref{subp15b}) is independent of $K_{\text{max}}$ -- it cancels between
${\rm SUM_{IR}(QCED)}$ and $D_\rQCED$. Having resummed these terms, we generate, in order to maintain exactness order by order in perturbation theory in both 
$\alpha$ and $\alpha_s$, the residuals $\tilde{\bar\beta}_{n,m}$ -- the latter
are computed iteratively to match the attendant 
exact results to all orders in $\alpha$ and $\alpha_s$ as explained
in Refs.~\cite{qced,irdglap1,irdglap2}.\par
We note that, as it is explained in Refs.~\cite{qced}, 
the new non-Abelian residuals $\tilde{\bar\beta}_{m,n}$ 
allow rigorous shower/ME matching via their shower subtracted analogs:
\begin{equation}
\tilde{\bar\beta}_{n,m}\rightarrow \hat{\tilde{\bar\beta}}_{n,m}
\end{equation}
where the $\hat{\tilde{\bar\beta}}_{n,m}$ have had all effects in the showers
associated to the $\{F_j\}$ removed from them.
The connection with the differential distributions in MC@NLO
can be seen as follows. The MC@NLO differential cross section 
can be represented~\cite{mcatnlo} as follows:
\begin{equation}
\begin{split}
d\sigma_{MC@NLO}&=\left[B+V+\int(R_{MC}-C)d\Phi_R\right]d\Phi_B[\Delta_{MC}(0)+\int(R_{MC}/B)\Delta_{MC}(k_T)d\Phi_R]\nonumber\\
&\qquad\qquad +(R-R_{MC})\Delta_{MC}(k_T)d\Phi_Bd\Phi_R
\end{split}
\end{equation}
where $B$ is Born distribution, $V$ is the regularized virtual contribution,
$C$ is the corresponding counter-term required at exact NLO, $R$ is the respective
exact real emission distribution for exact NLO, $R_{MC}=R_{MC}(P_{AB})$ is the parton shower real emission distribution
so that the Sudakov form factor is 
$$\Delta_{MC}(p_T)=e^{[-\int d\Phi_R \frac{R_{MC}(\Phi_B,\Phi_R)}{B}\theta(k_T(\Phi_B,\Phi_R)-p_T)]}$$,
where as usual it describes the respective no-emission probability. The respective Born and real emission differential phase spaces are denoted by $d\Phi_A, \; A=B,\; R$, respectively.
From comparison with (\ref{subp15b}) restricted to its QCD aspect we get the identifications, accurate to ${\cal O}(\alpha_s)$,
\begin{equation}
\begin{split}
\frac{1}{2}\hat{\tilde{\bar\beta}}_{0,0}&= \bar{B}+(\bar{B}/\Delta_{MC}(0))\int(R_{MC}/B)\Delta_{MC}(k_T)d\Phi_R\\
\frac{1}{2}\hat{\tilde{\bar\beta}}_{1,0}&= R-R_{MC}-B\tilde{S}_{QCD}
\label{eq-mcnlo}
\end{split}
\end{equation}
where we defined~\cite{mcatnlo} $$\bar{B}=B(1-2\alpha_s\Re{B_{QCD}})+V+\int(R_{MC}-C)d\Phi_R$$ and we understand here
that the DGLAP-CS kernels in $R_{MC}$ are to be taken as the IR-improved ones
as we exhibit below~\cite{irdglap1,irdglap2}. Here we have introduced the QCD virtual and real infrared functions
$B_{QCD}$ and $\tilde{S}_{QCD}$ respectively given in Refs.~\cite{irdglap1,irdglap2} which are understood to be DGLAP-CS synthesized as explained in Refs.~\cite{qced,irdglap1,irdglap2} to avoid doubling counting of effects.
The way to the extension of frameworks such as MC@NLO to exact higher
orders in $\{\alpha_s,\;\alpha\}$ is therefore open via our $\hat{\tilde{\bar\beta}}_{n,m}$
and will be taken up elsewhere~\cite{elswh}.
\par
We stress that in Refs.~\cite{irdglap1,irdglap2,herwiri} the methods we employ for resummation of the QCD theory have been shown to be
fully consistent with the methods in Refs.~\cite{stercattrent1,scet1}. 
What is shown in Refs.~\cite{irdglap1,irdglap2,herwiri}
is that the methods in Refs.~\cite{stercattrent1,scet1}
give approximations to our hard gluon residuals  $\hat{\tilde{\bar\beta}}_{n}$;
 for, the methods in Refs.~\cite{stercattrent1,scet1}, unlike the master formula in (\ref{subp15b}), are not exact results. Specifically, the threshold-resummation
methods in Refs.~\cite{stercattrent1}, using the result
that, for any function $f(z)$,
$$\left|\int_0^1 dz z^{n-1}f(z)\right|\le(\frac{1}{n})\max_{z\in [0,1]} {|f(z)|},$$
drop non-singular contributions to the cross section at $z\rightarrow 1$
in resumming the logs in $n$-Mellin space. The SCET theory in Refs.~\cite{scet1}
drops terms of ${\cal O}(\lambda)$ at the level of the amplitude, where $\lambda=\sqrt{\Lambda/Q}$ for a process with the hard scale $Q$ with $\Lambda \sim .3\text{GeV}$ so that, for $Q\sim 100\text{GeV}$, we have $\lambda\cong 5.5\%$.
From the known equivalence of the two approaches, the errors in the threshold resummation must be similar. Evidently, we can only use these approaches as a guide to our  new non-Abelian residuals as we develop results for the sub-1\% precision regime.\par
The discussions just completed naturally bring us to the attendant evolution of the $\{F_j\}$; for, 
in order to have a strict control on the theoretical precision
in (\ref{bscfrla}), we need both the resummation of the reduced cross section
and that of the latter evolution.
\par 
When the QCD restriction of the formula in (\ref{subp15b}) is applied to the
calculation of the kernels, $P_{AB}$, in the DGLAP-CS theory itself, 
we get an improvement
of the IR limit of these kernels, an IR-improved DGLAP-CS theory~\cite{irdglap1,irdglap2} in which large IR effects are resummed for the kernels themselves.
The resulting new resummed kernels, $P^{\exp}_{AB}$ are given in Refs.~\cite{irdglap1,irdglap2,herwiri} and are reproduced here for completeness:
{\small
\begin{align}
P^{\exp}_{qq}(z)&= C_F \FYFS(\gamma_q)e^{\frac{1}{2}\delta_q}\left[\frac{1+z^2}{1-z}(1-z)^{\gamma_q} -f_q(\gamma_q)\delta(1-z)\right],\nonumber\\
P^{\exp}_{Gq}(z)&= C_F \FYFS(\gamma_q)e^{\frac{1}{2}\delta_q}\frac{1+(1-z)^2}{z} z^{\gamma_q},\nonumber\\
P^{\exp}_{GG}(z)&= 2C_G \FYFS(\gamma_G)e^{\frac{1}{2}\delta_G}\{ \frac{1-z}{z}z^{\gamma_G}+\frac{z}{1-z}(1-z)^{\gamma_G}\nonumber\\
&\qquad +\frac{1}{2}(z^{1+\gamma_G}(1-z)+z(1-z)^{1+\gamma_G}) - f_G(\gamma_G) \delta(1-z)\},\nonumber\\
P^{\exp}_{qG}(z)&= \FYFS(\gamma_G)e^{\frac{1}{2}\delta_G}\frac{1}{2}\{ z^2(1-z)^{\gamma_G}+(1-z)^2z^{\gamma_G}\},
\label{dglap19}
\end{align}}
where the superscript ``$\exp$'' indicates that the kernel has been resummed as
predicted by Eq.\ (\ref{subp15b}) when it is restricted to QCD alone, where
the YFS~\cite{yfs} infrared factor 
is given by $\FYFS(a)=e^{-C_Ea}/\Gamma(1+a)$ where $C_E$ is Euler's constant
and where we refer the reader to Refs.~\cite{irdglap1,irdglap2} for the
detailed definitions of the respective resummation functions $\gamma_A,\delta_A,f_A, A=q,G$
\footnote{The improvement in Eq.\ (\ref{dglap19}) 
should be distinguished from the 
resummation in parton density evolution for the ``$z\rightarrow 0$'' 
Regge regime -- see for example Refs.~\cite{ermlv,guido}. This
latter improvement must also be taken into account 
for precision LHC predictions.}. $C_F$($C_G$) is the quadratic Casimir invariant for the quark(gluon) color representation respectively.
These new kernels 
yield a new resummed scheme for the parton density functions (PDF's) and the reduced cross section: 
\begin{equation}
\begin{split}
F_j,\; \hat\sigma &\rightarrow F'_j,\; \hat\sigma'\; \text{for}\\
P_{Gq}(z)&\rightarrow P^{\exp}_{Gq}(z), \text{etc.},
\end{split}
\label{newscheme1}
\end{equation}
with the same value for $\sigma$ in (\ref{bscfrla}) with improved MC stability
as discussed in Refs.~\cite{herwiri} -- there is no need 
for an IR cut-off `$k_0$'
parameter in the attendant parton shower MC based on the new kernels.
We point-out that, while the degrees of freedom
below the IR cut-offs in the usual showers are dropped in those showers,
in the showers in HERWIRI1.031,
as one can see from (\ref{subp15b}), these degrees of freedom are integrated over and included in the calculation in the process of generating the Gribov-Lipatov exponents $\gamma_A$ in (\ref{dglap19}). We note also that the new kernels
agree with the usual kernels at ${\cal O}(\alpha_s)$ as the differences between them start in ${\cal O}(\alpha_s^2)$. This means that the NLO matching formulas
in the MC@NLO and POWHEG frameworks apply directly 
to the new kernels for exact
NLO ME/shower matching. 
\par
For completeness, we feature in Fig.~1 the basic physical idea underlying the new kernels as it was already discussed by Bloch and Nordsieck~\cite{bn1}: 
\begin{figure}[h]
\begin{center}
\epsfig{file=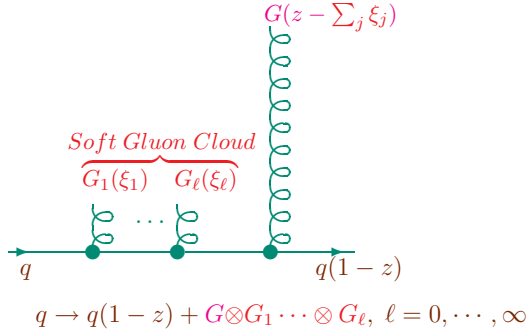,width=70mm}
\end{center}
\label{fig-bn-1}
\caption{Bloch-Nordsieck soft quanta for an accelerated charge.}
\end{figure}
an accelerated charge generates a coherent state of very soft massless quanta of the respective gauge field so that one cannot know which of the infinity of possible states
one has made in the splitting process $q(1)\rightarrow q(1-z)+G\otimes G_1\cdots\otimes G_\ell,\; \ell=0,\cdots,\infty$ illustrated in Fig.~1.
The new kernels take this effect into account by resumming the 
terms ${\cal O}\left((\alpha_s \ln(\frac{q^2}{\Lambda^2})\ln(1-z))^n\right)$
when $z\rightarrow 1$ is the IR limit. As one can see in (\ref{newscheme1}) and (\ref{bscfrla}), when the usual kernels are used these terms
are generated order-by-order in the solution for the cross section
$\sigma$ in (\ref{bscfrla}) and our 
resumming them enhances the convergence of the 
representation in (\ref{bscfrla}) for a given order of exactness in the
input perturbative components therein.  We now turn to
the illustration of this last remark in the context of the comparison of 
NLO parton shower/matrix element matched predictions to recent LHC data.\par

\section{Interplay of NLO Shower/ME Precision and IR-Improved DGLAP-CS Theory}

The new MC HERWIRI1.031~\cite{herwiri} gives the first realization of the new IR-improved kernels in the HERWIG6.5~\cite{herwig} environment. Here, 
we compare it with HERWIG6.510, both with and without
the MC@NLO~\cite{mcatnlo} exact ${\cal O}(\alpha_s)$ correction
to illustrate the interplay between the attendant precision in NLO ME matched parton shower MC's  
and the new IR-improvement for the kernels where we use the new LHC data for our baseline for the comparison.\par
\begin{figure}[h]
\begin{center}
\setlength{\unitlength}{0.1mm}
\begin{picture}(1600, 930)
\put( 370, 770){\makebox(0,0)[cb]{\bf (a)} }
\put(1240, 770){\makebox(0,0)[cb]{\bf (b)} }
\put(   -50, 0){\makebox(0,0)[lb]{\includegraphics[width=80mm]{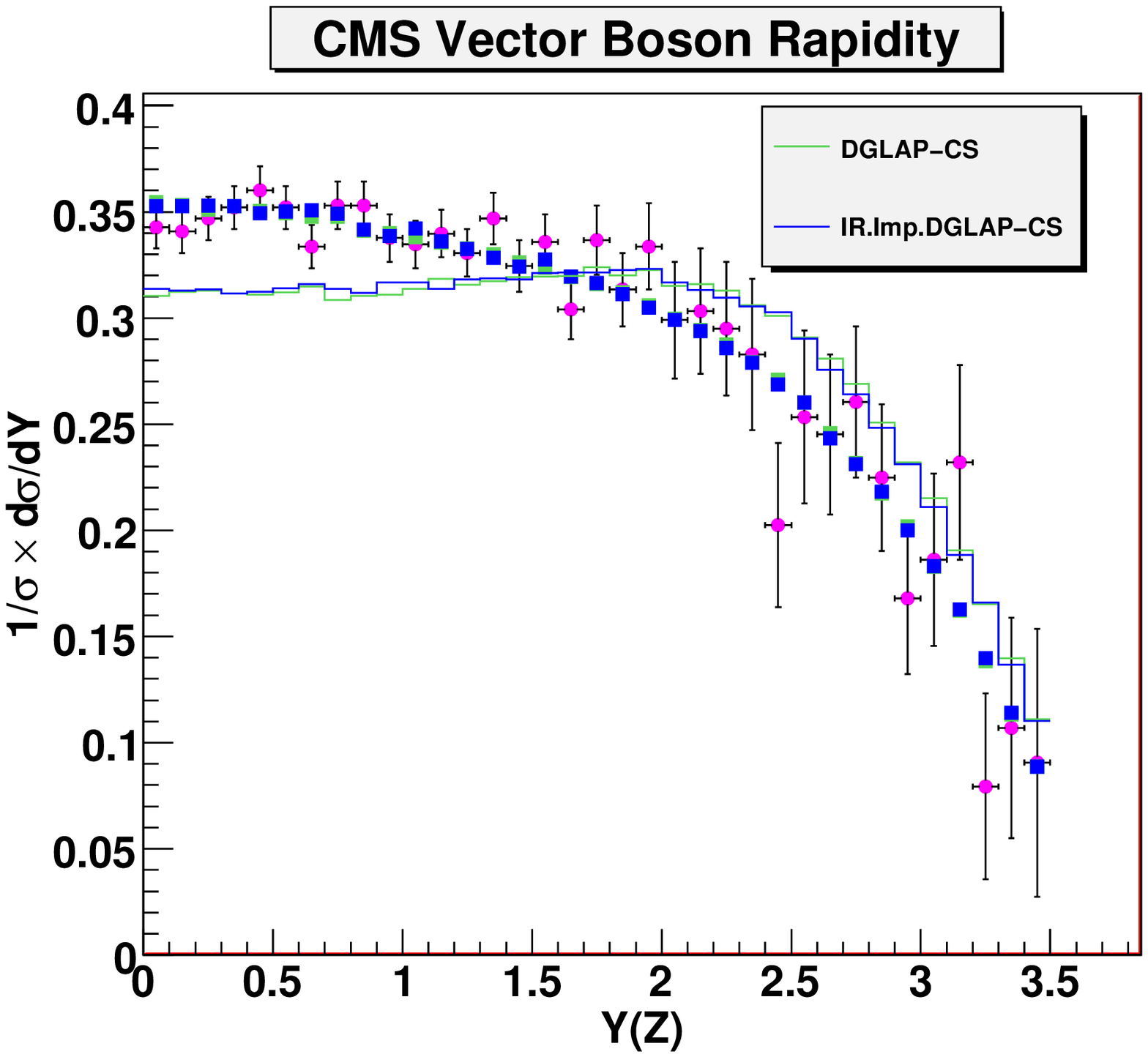}}}
\put( 830, 0){\makebox(0,0)[lb]{\includegraphics[width=80mm]{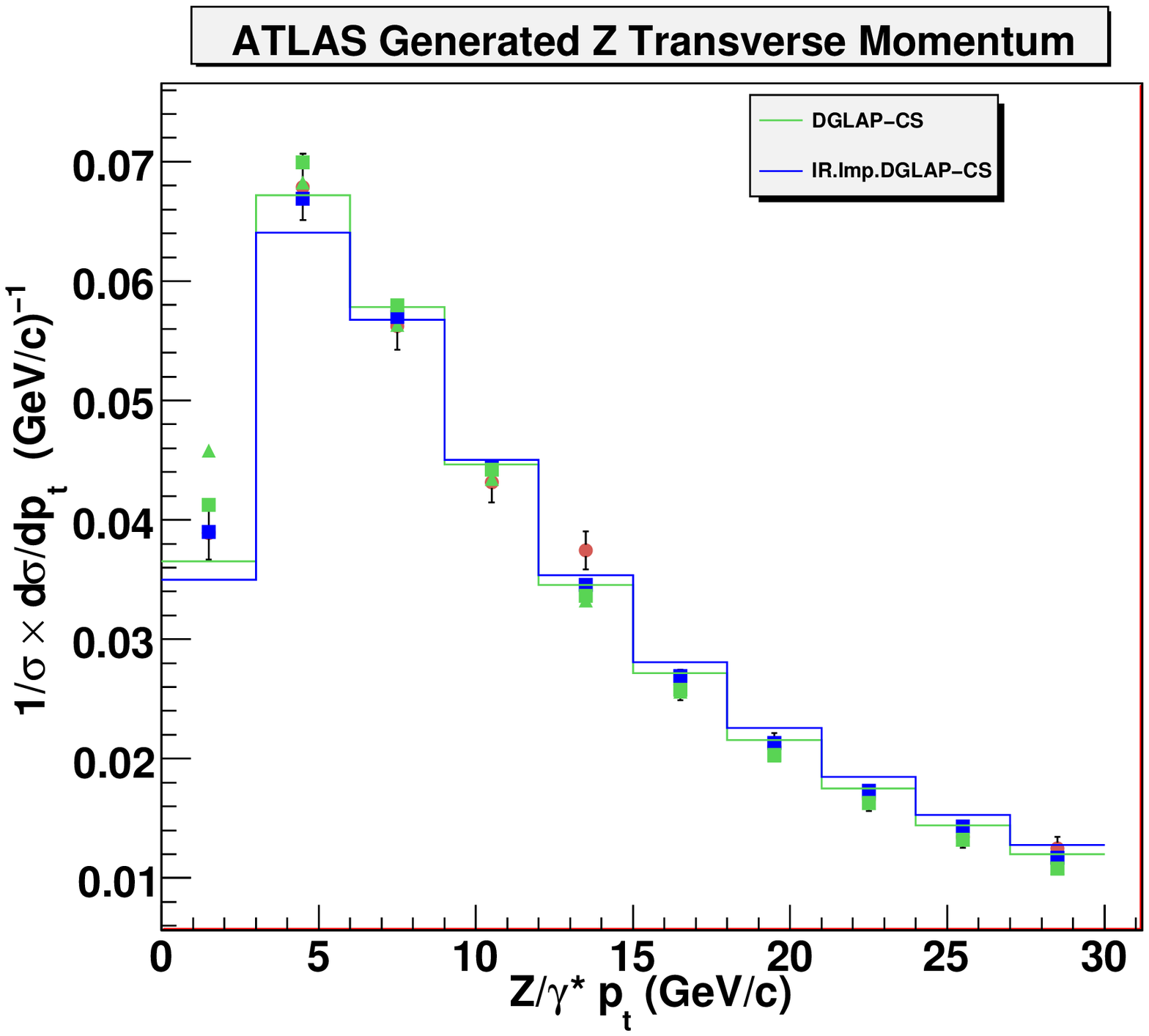}}}
\end{picture}
\end{center}
\caption{\baselineskip=8pt Comparison with LHC data: (a), CMS rapidity data on
($Z/\gamma^*$) production to $e^+e^-,\;\mu^+\mu^-$ pairs, the circular dots are the data, the green(blue) lines are HERWIG6.510(HERWIRI1.031); 
(b), ATLAS $p_T$ spectrum data on ($Z/\gamma^*$) production to (bare) $e^+e^-$ pairs,
the circular dots are the data, the blue(green) lines are HERWIRI1.031(HERWIG6.510). In both (a) and (b) the blue(green) squares are MC@NLO/HERWIRI1.031(HERWIG6.510($\rm{PTRMS}=2.2$GeV)). In (b), the green triangles are MC@NLO/HERWIG6.510($\rm{PTRMS}=$0). These are otherwise untuned theoretical results. 
}
\label{fig2-nlo-iri}
\end{figure}

More precisely, for the single $Z/\gamma*$ production at the LHC, we show in Fig.~\ref{fig2-nlo-iri} in panel (a) the comparison between the MC predictions and the CMS rapidity data~\cite{cmsrap} and in panel
(b) the analogous comparison with the ATLAS $P_T$ data, where the rapidity data  are the combined $e^+e^--\mu^-\mu^+$ results and the $p_T$ data are those for the bare $e^+e^-$ case, as these are the data that correspond to the theoretical
framework of our simulations -- we do not as yet have complete realization of all the corrections involved in the other ATLAS data in Ref.~\cite{atlaspt}. 
These results should be viewed from the perspective of our analysis in Ref.~\cite{herwiri} of the FNAL data on the single $Z/\gamma^*$ production in 
$\text{p}\bar{\text{p}}$ collisions at 1.96 TeV.\par
Specifically, in Fig.~11 of the second paper in Ref.~\cite{herwiri}, we showed that, when the intrinsic rms $p_T$ parameter $\rm{PTRMS}$ is set to 0 in HERWIG6.5, the simulations for MC@NLO/HERWIG6.510 give a good fit to the CDF rapidity distribution data~\cite{galea} therein but they do not give a satisfactory fit to the D0 $p_T$ distribution data~\cite{d0pt} therein whereas the results for MC@NLO/HERWIRI1.031 give good fits to both sets of data with the $\rm{PTRMS} =0$. Here $\rm{PTRMS}$ corresponds to an intrinsic Gaussian distribution in $p_T$. The authors of HERWIG~\cite{mike2} have emphasized that to get good fits to both sets of data, one may set $\rm{PTRMS}\cong 2$ GeV. Thus, in analyzing the new LHC data, we have set $\rm{PTRMS}=2.2$GeV in our HERWIG6.510 simulations while we continue to set $PTRMS=0$ in our HERWIRI simulations.
\par
Turning now with this perspective to the results in Fig.~\ref{fig2-nlo-iri}, we see a confirmation of the finding of the HERWIG authors. To get a good fit to both the CMS rapidity data and the ATLAS $p_T$ data, one needs to set $\rm{PTRMS}\cong 2 \text{GeV}$~\cite{skands} in the MC@NLO/HERWIG6510 simulations. We again see that at LHC one gets a good fit to the data for both the rapidity and the $p_T$ spectra in the MC@NLO/HERWIRI1.031 simulations with $\rm{PTRMS}=0$. In quantitative terms, the $\chi^2/\text{d.o.f.}$ for the rapidity data and $p_T$ data are (.72,.72)((.70,1.37)) for the 
MC@NLO/HERWIRI1.031(MC@NLO/HERWIG
6510($\rm{PTRMS}$=2.2GeV)) simulations. For the
 MC@NLO/HERWIG6510($\rm{PTRMS}$=0) simulations the corresponding results are (.70,2.23).
\par 
Thus, we see that the usual DGLAP-CS kernels require the introduction of a {\em hard}
intrinsic Gaussian spread in $p_T$  inside the proton to reproduce the LHC data on the $p_T$ distribution of the 
$Z/\gamma^*$ in the pp collisions whereas the IR-improved kernels give in fact a better fit to the data without the introduction of such a hard intrinsic component to the motion of the proton's constituents. The {\em hardness}
of this  $\rm{PTRMS}$ is entirely ad hoc; it is in contradiction with the results of all successful models of the proton wave-function~\cite{pwvfn},
wherein the scale of the corresponding $\rm{PTRMS}$ is found to be 
$\lesssim 0.4$GeV. More importantly, it contradicts the known experimental observation of precocious Bjorken scaling~\cite{scaling,bj1}, where the SLAC-MIT experiments show that Bjorken scaling occurs already at $Q^2=1_+$ GeV$^2$
for $Q^2=-q^2$ with q the 4-momentum transfer from the electron to the proton
in the famous deep inelastic electron-proton scattering process whereas, if the proton constituents really had a Gaussian intrinsic $p_T$ distribution with $\rm{PTRMS}\cong 2$GeV, these observations 
would not be possible. What can now say is that the ad hoc ``hardness''
of the $\rm{PTRMS}\cong 2.2$GeV value is really just a phenomenological representation of the more fundamental dynamics realized by the IR-improved DGLAP-CS theory. This raises the question of whether it is possible
to tell the difference between the two representations of the data in 
Fig.~\ref{fig2-nlo-iri}.\par
\begin{figure}[h]
\begin{center}
\epsfig{file=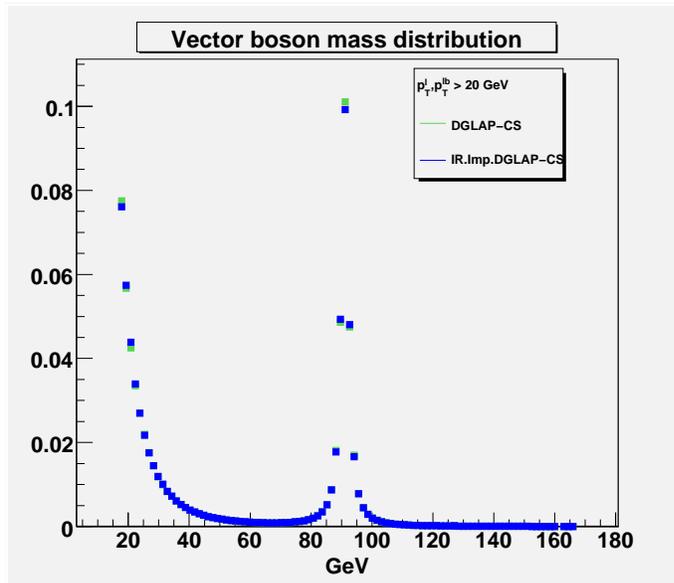,width=90mm}
\end{center}
\label{fig3-nlo-iri}
\caption{Normalized vector boson mass spectrum at the LHC for $p_T(\text{lepton}) >20$ GeV.}
\end{figure}

Physically, one expects that more detailed observations should be able to distinguish the two. Specifically, we show in Fig.~\ref{fig3-nlo-iri} the  MC@NLO/HERWIRI
1.031(blue squares) and MC@NLO/HERWIG6510($\rm{PTRMS}$=2.2GeV) (green squares) predictions
for the $Z/\gamma^*$ mass spectrum when the decay lepton pairs are required to 
satisfy the LHC type requirement that their transverse momenta $\{p^\ell_T, p^{\bar\ell}_T\}$ exceed $20$ GeV.
We see that the high precision data such as the LHC ATLAS and CMS
experiments will have
(each already has over $5\times 10^6$ lepton pairs) would allow one to distinguish between the two sets
of theoretical predictions, as the peaks differ by 2.2\% for example.\par
Continuing in this way, we make a more detailed snap-shot of the region below
$10.0$GeV in Fig.~\ref{fig2-nlo-iri} (b) in which we plot the three featured theory predictions with finer binning, $0.5$GeV instead of $3.0$GeV. This 
is shown in Fig.~4. 
\begin{figure}[h]
\begin{center}
\epsfig{file=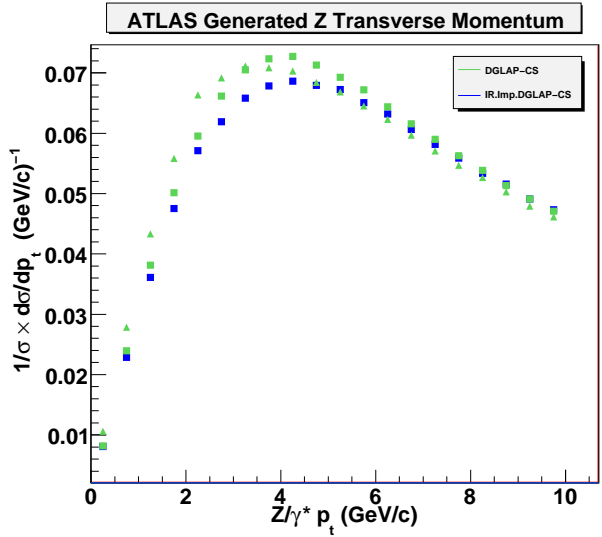,width=90mm}
\end{center}
\label{fig4-nlo-iri}
\caption{Normalized vector boson $p_T$ spectrum at the LHC 
for the ATLAS cuts as exhibited in Fig.~\protect{\ref{fig2-nlo-iri}} 
for the same conventions on the notation for the theoretical results
with the vector boson $p_T < 10$ GeV to illustrate the differences between
the three predictions.}
\end{figure}
We see that there are significant differences in the shapes of the three predictions that are testable
with the precise data that will be available
to ATLAS and CMS experiments.
Other such detailed observations
may also reveal 
the differences between the two descriptions of parton shower physics
and we will pursue these elsewhere~\cite{elswh}.
We await the release of the entire data sets from ATLAS and CMS.\par
\section{Conclusions}
We have shown that the realization of IR-improved DGLAP-CS theory 
in HERWIRI1.031, when used in the MC@NLO/HERWIRI1.031 exact ${\cal O}(\alpha_s)$ ME matched parton shower framework,
affords one the opportunity to explain, on an event-by-event basis, both the rapidity and the $p_T$ spectra of the $Z/\gamma^*$ in pp collisions
in the recent LHC data from CMS and ATLAS, respectively, without the need of an
unexpectedly hard intrinsic Gaussian $p_T$ distribution with rms value of $\rm{PTRMS}\cong 2$ GeV in the proton's wave function. We argue that this can be interpreted as providing a rigorous basis for the phenomenological correctness 
of such unexpectedly hard distributions insofar as describing these data using the usual unimproved DGLAP-CS showers is concerned and we have proposed 
that comparison of other distributions such as the invariant mass distribution 
with the appropriate cuts and the more detailed $Z/\gamma^*$ $p_T$ 
spectra in the regime below $10.0$GeV be used to
differentiate between the fundamental description of the parton shower physics in MC@NLO/HERWIRI1.031 and these phenomenological representations
in MC@NLO/HERWIG6510. We have emphasized that the precociousness of Bjorken scaling argues against the fundamental correctness 
of the hard scale intrinsic $p_T$ ansatz with the unexpectedly hard value of $\rm{PTRMS}\cong 2$ GeV, as do the successful models~\cite{pwvfn} of the proton's wave function,
which would predict this value to be $\lesssim 0.4$GeV. We have the added bonus that the fundamental description in MC@NLO/HERWIRI1.031 can be systematically improved to the NNLO parton shower/ME matched level which we anticipate is a key ingredient in achieving the sub-1\% precision tag for such processes as single heavy gauge boson production at the LHC. Evidently, the use of ad hoc hard
scales in models would compromise
any discussion of the theoretical precision relative to what one could achieve
from the fundamental representation of the corresponding physics via IR-improved DGLAP-CS theory as it is realized in HERWIRI1.031 when employed in
MC@NLO/HERWIRI1.031 simulations. We are pursuing additional cross checks
of the latter simulations against the LHC data.
\par
In closing, two of us (A.M. and B.F.L.W.)
thank Prof. Ignatios Antoniadis for the support and kind 
hospitality of the CERN TH Unit while part of this work was completed.\par

\end{document}